# Invoking forbidden modes in $SnO_2$ nanoparticles using tip enhanced Raman spectroscopy

Venkataramana Bonu,* A. Das, A. K. Sivadasan, A. K. Tyagi, Sandip Dhara*

*Surface and Nanoscience Division, Indira Gandhi Center for Atomic Research, Kalpakkam-603102, India.*

Raman forbidden modes and surface defect related Raman features in $SnO_2$ nanostructures carry information about disorder and surface defects which strongly influence important technological applications like catalysis and sensing. Due to the weak intensities of these peaks, it is difficult to identify these features by using conventional Raman spectroscopy. Tip enhanced Raman spectroscopy (TERS) studies conducted on $SnO_2$ nanoparticles (NPs) of size 4 and 25 nm have offered significant insights of prevalent defects and disorders. Along with one order enhancement in symmetry allowed Raman modes, new peaks related to disorder and surface defects of $SnO_2$ NPs were found with significant intensity. Temperature dependent Raman studies were also carried out for these NPs and correlated with the TERS spectra. For quasi-quantum dot sized 4 nm NPs, the TERS study was found to be the best technique to probe the finite size related Raman forbidden modes.

Correspondence to: ramana9hcu@gmail.com, dhara@igcar.gov.in

**Introduction**

SnO$_2$ is an *n*-type wide band gap semiconductor with direct band gap of 3.6 eV. It is widely used as gas sensor,[1-3] transparent conductors (when doped suitably),[1,4] solar cells,[1,5] Li-ion batteries and also as a catalyst.[1,6] The surface chemistry of a SnO$_2$ structures is a crucial parameter in all these applications.[1-3] High surface to volume ratio in nanostructured SnO$_2$ extensively manipulates its physiochemical properties, and its applications.[1-6] In our earlier report we have studied the role of in-plane and bridging 'O' vacancy related surface defects in the detection of CH$_4$, a green house gas.[2] Raman forbidden modes of $E_{u}$(TO), $A_{2u}$(TO), $B_{1u}$, and $A_{2u}$(LO) at 248, 502, 544, 694 cm$^{-1}$, respectively and surface defect related vibrational features at 400-700 cm$^{-1}$ were observed in the Raman spectra of nanostructured SnO$_2$ along with symmetry allowed Raman modes in SnO$_2$.[2,7-11] Raman forbidden modes appear due to the breakdown of the selection rule as finite size effect in these nanostructures.[2,7-11] Importantly these modes carry the information about the surface defects and disorder in the material. Rumyantseva *et al*[2] explicitly reported intensity of surface defect related vibrational features at 400-700 cm$^{-1}$ to be proportional to CO response whereas it was inversely proportional to NO$_2$ response. So, exploring the surface properties of SnO$_2$ is a technologically important task. It is not always easy to identify presence of all these Raman forbidden modes and surface defect related vibrational features by using normal Raman spectroscopy due to their weak intensity. Particularly in quantum and quasi-quantum sized NPs, these features are overlapped with the 'O' vacancy related broad defect band at 570 cm$^{-1}$.[8,12]

Tip enhanced Raman spectroscopy (TERS) is a non destructive technique and is known to detect even a single molecule using the evanescent field of noble metal coated surface probe microscopic (SPM) tip.[13] Defect D-band of a single walled carbon nanotube (SWNT),[14] strain in the Si on insulator (SOI)[15] and compositional inhomogeneity and phase instability of GaN/InGaN quantum well[16] were studied by the TERS technique. Fazio *et al*[17] studied the surface enhanced Raman spectroscopy (SERS) on bulk SnO$_2$, thin film and colloidal solution. A new peak at 600 cm$^{-1}$ was explained invoking the gradient field Raman mechanism and correlated to IR active $A_{2u}$ mode. However, they have not observed any considerable enhancement of intensity in the whole spectrum due to SERS. To the best of our knowledge there is hardly any TERS study on SnO$_2$ nanostructures to explore modes either due to finite size effect or related to surface defects.

Here we report the Raman forbidden modes and surface defect related Raman features in $SnO_2$ nanoparticles (NPs) using TERS spectroscopy. $SnO_2$ NPs with large diameter of 25 nm (bulk like properties) and quasi-quantum confinement size of 4 nm with pronounced surface defects were used for the present study. Thermally introduced defects in the temperature dependent Raman spectroscopic study are correlated with the TERS studies.

**Experimental**

$SnO_2$ NPs were synthesized by annealing pristine quantum dots of size 2.4 nm at 300 and 800 $^oC$ in horizontal quartz tube furnace in presence of air atmosphere for 1hr. The pristine sample was synthesized by soft chemical method and a detailed synthesis and characterization of these NPs was reported elsewhere.[3] Morphological studies of the synthesized materials were performed by high resolution transmission electron microscopy (HRTEM; Zeiss Libra 200) and field emission scanning electron microscopy (FESEM; Zeiss SUPRA 55) for samples annealed at different temperatures. For TERS studies the $SnO_2$ NPs were dispersed in the isopropanol and then drop casted on to the Cu coated Si substrate. Substrate was then dried at 100 $^oC$ to evaporate the isopropanol. TERS study was performed in micro-Raman spectrometer (InVia; Renishaw) coupled with scanning probe microscopy (Nanonics MultiView 4000). Micro-Raman spectroscopy (InVia; Renishaw) using 514.5 nm excitation of $Ar^+$ laser was carried out with 1800 gr.mm$^{-1}$ grating and thermoelectric cooled CCD detector in the backscattering configuration. For TERS studies, specially designed high dielectric contrast and Raman background free 200 nm Au nanoparticle cantilevered glass probe (Nanonics) was engaged as TERS probe in tapping mode (Fig. 1a). Morphology of the NPs cluster annealed at 800 $^oC$ (Fig. 1b) on which TERS studies were performed was recorded by the atomic force microscopy (AFM, Nanonics MultiView 4000). Temperature dependent micro-Raman measurements were performed in adiabatic stage (Linkam, UK). The TERS and temperature dependent Raman experiments were conducted on several positions of the dispersed material to ensure the reproducibility.

**Results and Discussion**

Rutile tetragonal $SnO_2$, containing two Sn and four O atoms in a single unit cell, belongs to the space group $D_{4h}^{14}$. According to the group theory, the representation of normal vibration modes at the center of the Brillion zone is $\Gamma = A_{1g} + A_{2g} + 2A_{2u} + B_{1g} + B_{2g} + 2B_{1u} + E_g + 4E_u$. Among these, three non-degenerate modes $B_{1g}$, $B_{2g}$ and $A_{1g}$ (vibrating in the plane perpendicular to the $c$-axis) and another doubly degenerate $E_g$ mode (vibrating in the direction of the $c$-axis) are Raman active. The singlet $A_{2u}$ and the triply degenerate $E_u$ modes are infra-red (IR) active. Modes corresponding to the $A_{2g}$ and $B_{1u}$ symmetries are silent.[8] Corresponding to the $E_u$ mode, both Sn and O atoms vibrate in the plane perpendicular to the $c$-axis. The silent modes correspond to vibrations of the Sn and O atoms along the $c$-axis ($B_{1u}$). Fig. 2 shows the Raman spectra of the NPs annealed at 800 °C with and without TERS tip. Inset in Fig. 2 shows the FESEM image of the NPs. Average size of these spherical NPs was found to be ~25 nm. AFM image in Fig. 1b illustrates the topography of the 25 nm NPs cluster. TERS studies were performed on this cluster. Only two symmetry allowed Raman modes at 633 ($A_{1g}$) and 775 cm$^{-1}$ ($B_{2g}$) were observed for the 25 nm NPs without the use of TERS tip (Fig. 2). Intensity of the Raman spectrum recorded in the presence of TERS tip was enhanced by one order with respect to the spectrum recorded without the tip (Fig. 2). In addition to this enhancement, a peak at 477 cm$^{-1}$ corresponding to symmetry allowed $E_g$ mode was observed. In addition, new peaks appeared at 248, 502, and 694 cm$^{-1}$ were assigned to IR active $E_u$(TO), $A_{2u}$(TO), and $A_{2u}$(LO) modes, respectively along with the forbidden Raman mode $B_{1u}$ at 544 cm$^{-1}$.[2,7-11] Abello et al.[7] have proposed that the relaxation of the $k$=0 selection rule is progressive when the rate of disorder increases or size decreases, and IR modes can become weakly active when the structural changes induced by disorder and size effects take place. The intensity in the region 400-800 cm$^{-1}$ was also found to enhance (Fig. 2) in the TERS measurement. Thus it is obvious from Fig. 2 that TERS study enables observation of these obscured peaks in the sample.

In order to study the temperature effect on disorder and surface defect related modes and also for further comparison with TERS measurement, we conducted temperature dependent Raman study on the 25 nm NPs by varying the temperature from 80 K to 773 K (Fig. 3). Unlike TERS measurement, enough quantity of material was used for this study. Spectrum at room temperature (RT) showed a weak signature of $A_{2u}$(LO) and surface band at 400-800 cm$^{-1}$ (Fig. 3). Raman spectrum at 80 K showed highest intensity for symmetry allowed Raman modes, whereas forbidden modes were not observed at this temperature (Fig. 3). It indicates that the disorder of the material minimizes at low temperature. While increasing the temperature, intensity of the modes related to the disorder and surface defect related features were also increased (Fig. 3). At high temperatures of 623 and 773 K, IR active modes at $A_{2u}$(TO), and $A_{2u}$(LO) were found to appear with noticeable intensities. However, the intensities corresponding to IR active $E_u$(TO), and forbidden $B_{1u}$ peaks are not as clear as seen in TERS measurement (Fig. 2). With increase in temperature, intensity of symmetry allowed Raman modes $A_{1g}$ and $B_{2g}$ (Vibrates perpendicular to the $c$-axis) were found to decrease until a temperature of 573 K, and then further increased for the temperatures of 623 and 773 K. Whereas the peak $E_g$ (vibrates along the *c*-axis) gradually disappeared while increasing the temperature (Fig. 3). Lowering in the Raman intensity may be a result of the increase in bond length, where the peak broadening as well decrease in intensity took place.[18,19] Another reason for this effect may be because of the fact that the population of particles in the ground state decreases with increasing temperature resulting in lowering of stokes lines intensity. Intensity of the peaks $A_{1g}$ and $B_{2g}$ again increased for the temperature of 623 and 773 K (Fig. 3). In our earlier report we have demonstrated the photoluminescence related to –OH groups (which forms a defect energy state) in different size $SnO_2$ NPs including the 25 nm NPs using an excitation of 514.5 nm.[12] Xie *et al.*[20] reported that the optical absorption of the laser due to defects in the metal oxides will cause the reduction in the Raman signal intensity. In the present case the defect energy state formed due to –OH groups might be absorbing some percentage of the laser (514.5 nm). However at high temperatures these –OH groups will be removed, consequently there will be no optical absorption by –OH group related defects. This might be the reason for the enhancement in the intensity of the peaks $A_{1g}$ and $B_{2g}$ at high temperatures. The peak position corresponding to $A_{1g}$ varies in the range of 637-622 cm$^{-1}$ for increasing temperatures from 80 to 773K (Fig. 3). This red shift in the peak position is due to the increase of anharmonicities in the lattice with increasing temperature.[19] However, the forbidden Raman modes are not clear until a temperature of 573 K (Fig. 3). The appearance of forbidden Raman modes at high temperature is thus linked to the increase in the defects or disorder in the material. The red shift in $A_{1g}$ peak, in the TERS study of the 25 nm NPs, was found to be only 3 cm$^{-1}$ (Fig. 2). It indicates that the TERS technique due to surface plasmon effect of its tip enhances the intensity of the weak defect peaks without creating further disorder in the system.

Figure 4 shows the Raman spectra of the NPs annealed at 300 °C with and without TERS tip. Inset shows the HRTEM image of the NPs used for the TERS studies. Size of the NPs was measured to be ~4 nm. Symmetry allowed $A_{1g}$ at 632 cm$^{-1}$ and defect, $D$ mode at 570 cm$^{-1}$ were observed in the Raman spectrum without TERS tip. The $D$ mode at 570 cm$^{-1}$ appears only in small size (diameter around 9 nm and below) $SnO_2$ nanostructures.[8,12] Optically inactive $A_{2g}$ mode was reported to appear at a value of 575 cm$^{-1}$ following Matossi force constant model that considered modified bond length, space symmetry reduction and lattice distortion due to 'O' vacancies and local lattice disorders in $SnO_2$ NPs.[12] Along with a newly appeared weak symmetry allowed $B_{2g}$ mode at 775 cm$^{-1}$, three times enhancement of Raman spectral intensity was observed for 4 nm NPs in the TERS measurement with respect to the spectrum recorded without tip (Fig. 4). Additionally, the IR active $A_{2u}$(TO) and $A_{2u}$(LO) modes were distinctly observed in the TERS spectrum (Fig. 4). Raman spectra of the 4 nm NPs which were recorded at different temperatures from 80 - 573 K are shown in Fig. 4. Similar to the 25 nm NPs, the $A_{1g}$ peak intensity in 4 nm NPs was found to decrease and broaden with the increase in temperature. The *D* mode of defect origin was found to show a broadening effect as well as an improvement in intensity with increasing temperature. However, unlike the temperature dependent Raman spectra of 25 nm NPs, we do not observe any noticeable signature of the forbidden Raman modes for the 4 nm NPs (Fig. 5) This may be related to difficulty in creation of defects for reduced dimension $SnO_2$ NPs.[21] Presence of $A_{2u}$(TO) and $A_{2u}$(LO) is clear from TERS spectra of 4 nm NPs (Fig. 4). The micro Raman spectrum of 4 nm recorded at 573 K was then deconvoluted with Gaussian fittings to show peaks related to IR active $A_{2u}$(TO) and $A_{2u}$(LO) (Fig. 6). Notably authentic signature for the presence of IR active modes, $A_{2u}$(TO) and $A_{2u}$(LO), was distinctly provided in the TERS measurement (Fig. 4). These features are not clearly seen in micro Raman mode either at RT or at high temperature for the 4 nm NPs. This may be because of the overlapping with broad defect band *D* and another reason can be temperature induced broadness and red shift in these peaks. The localized (evanescent) field created by the TERS tip is more effective near surface of the NPs than that in the bulk of the sample due to the field gradient. Thus the strong field near the surface of NPs may be the reason for increased intensity corresponding to the defects related modes in case of measurements with TERS tip in comparison to the conventional Raman study.

**Conclusions**

Raman forbidden modes and surface defect related Raman features in 4 and 25 nm NPs are successfully probed by the tip enhanced Raman spectroscopy. Along with one order enhancement in symmetry allowed Raman modes, the obscured peaks related to disorder and surface defects of $SnO_2$ NPs are also observed in the TERS spectra of 25 nm NPs. The TERS spectrum of 4 nm NPs provides distinct signature of the Raman forbidden modes $A_{2u}$(TO) and $A_{2u}$(LO) which usually overlap with the 'O' vacancy related broad defect band *D* particularly in quantum and quasi-quantum $SnO_2$ NPs. Disorder and surface defect related bands in NPs are found to be subdued at low temperature of 80 K. This study is not only unravel interplay of surface defects as well disorder but also opens up ways to look into surface controlled reactions like sensing and catalytic effects in details.

**Acknowledgements**

We thank Mr. Avinash Patsha and Mr. Kishore K Madapu of SND/MSG/IGCAR for the valuable discussion.

**Figure Captions**

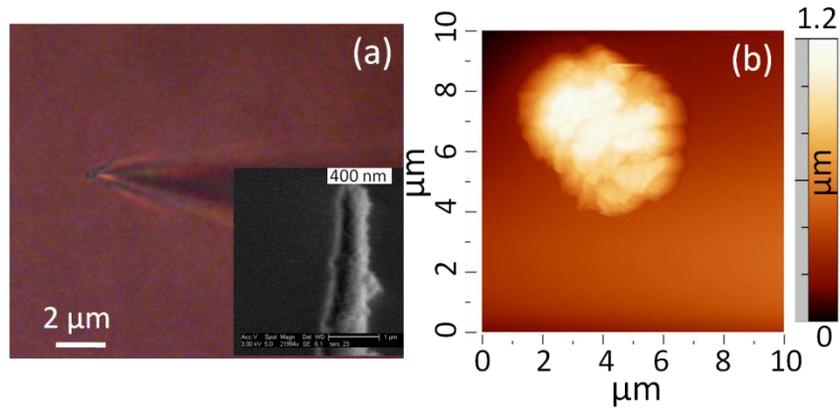

**Figure. 1** (a) Optical image of TERS tip with inset showing the SEM image and (a) AFM topological image of NPs annealed at 800 °C.

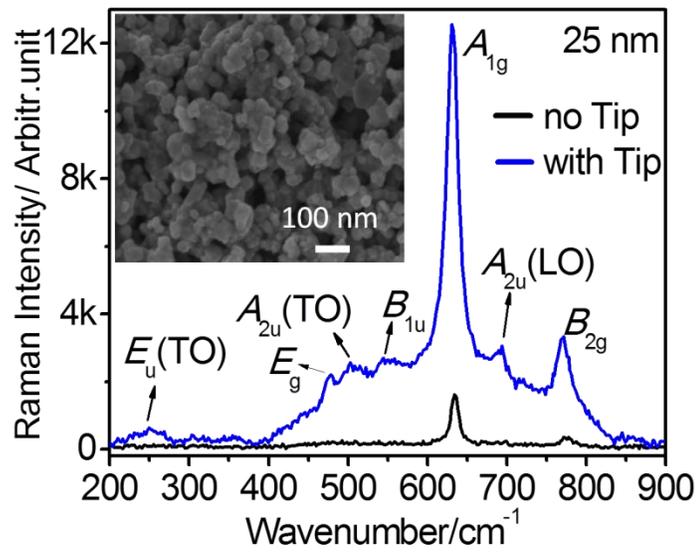

**Figure. 2** Raman spectra of $SnO_2$ NPs of size 25 nm in absence and presence of the TERS tip. Inset shows the FESEM image of the 25 nm NPs.

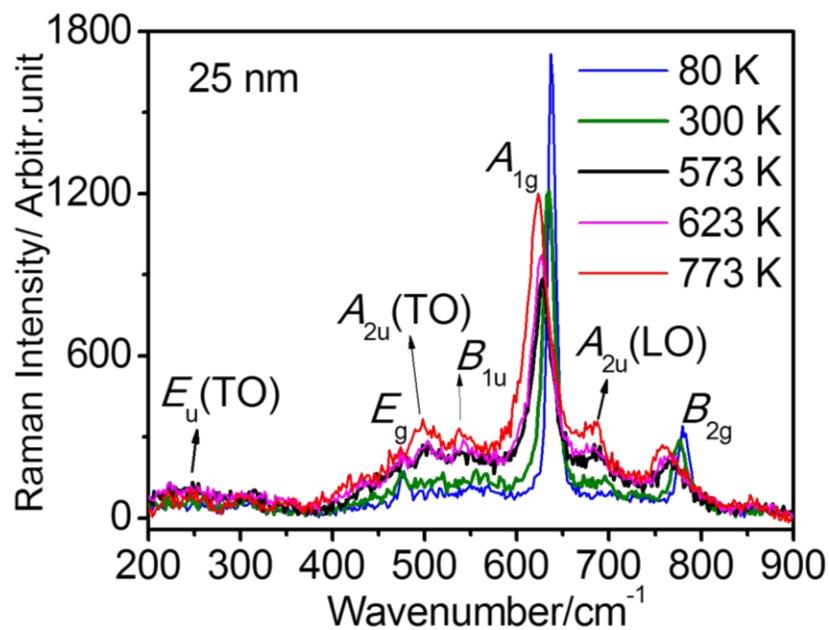

**Figure. 3** Temperature dependent Raman spectra of 25 nm NPs.

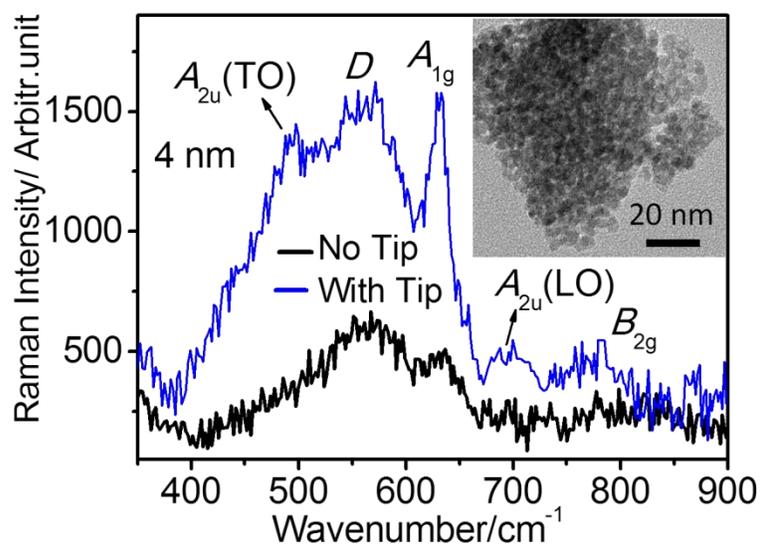

**Figure. 4** Raman spectra of $SnO_2$ NPs of size 4 nm in absence and presence of the TERS tip. Inset shows the HRTEM image of the 4 nm NPs.

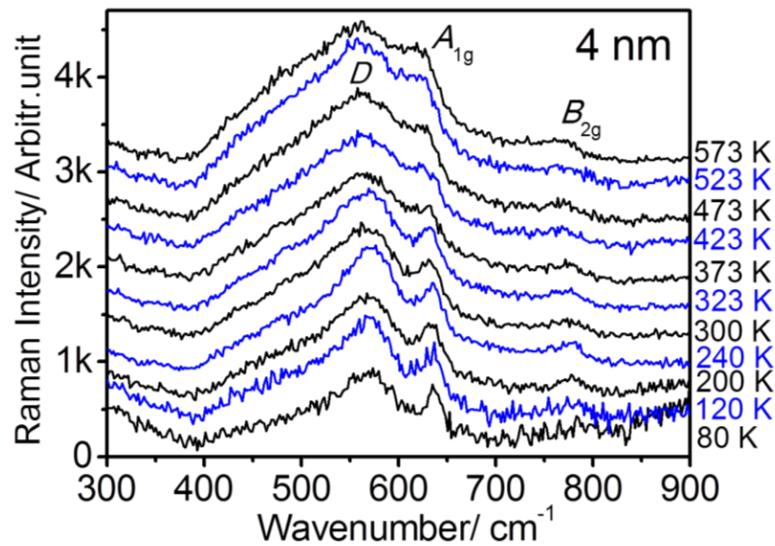

**Figure. 5** Temperature dependent Raman study of 4 nm NPs.

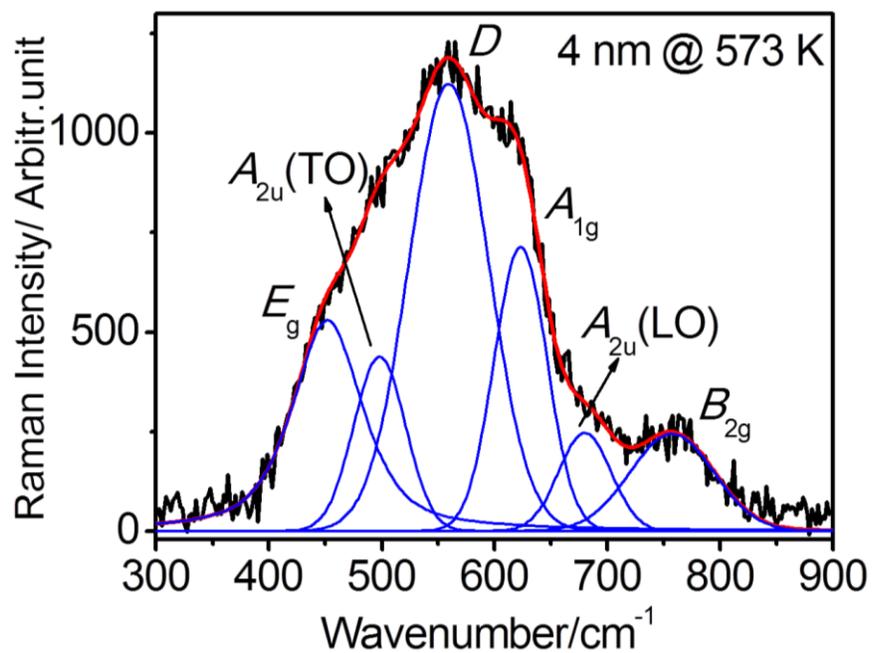

**Figure. 6** Gaussian deconvolution of Raman spectrum recorded at 573 K for 4 nm NPs.